\documentclass[fleqn,usenatbib]{mnras}
\usepackage{newtxtext,newtxmath}
\usepackage[T1]{fontenc}
\usepackage{ae,aecompl,graphicx,amsmath,comment,epsfig,psfrag}

\def\E33{\overline{E}_{33}}
\title[Constraining FRB properties]{A maximum likelihood
estimate of the parameters of the FRB population}

\author[Bhattacharyya et al.] {Siddhartha Bhattacharyya$^1$\thanks{siddhartha@phy.iitkgp.ac.in}, Himanshu Tiwari$^2$, Somnath Bharadwaj$^1$ and Suman Majumdar$^{2,3}$\\ 
$^1$Department  of Physics, Indian Institute of Technology, Kharagpur, India \\ 
$^2$Department of Astronomy, Astrophysics and Space Engineering, Indian Institute of Technology, Indore, India\\
$^3$Department of Physics, Blackett Laboratory, Imperial College, London SW7 2AZ, U.K.}
\begin{document}
\label{firstpage}
\pagerange{\pageref{firstpage}--\pageref{lastpage}}
\maketitle

\begin{abstract}
We consider a sample of $82$ non-repeating FRBs detected at Parkes, ASKAP, CHIME and UTMOST each of which operates over a different frequency range and has a different detection criteria. Using simulations, we perform a maximum likelihood analysis to determine the FRB population model which best fits this data. Our analysis shows that models  where the pulse scatter  broadening increases moderately  with redshift ($z$)  are  preferred over those  where this increases very sharply or  where scattering is absent. Further, models where the comoving event rate density is constant over $z$ are preferred over those  where it follows the cosmological star formation rate.  Two models for the host dispersion measure ($DM_{\rm host}$) distribution (a fixed and a random $DM_{\rm host}$) are found to  predict comparable results. We obtain the best fit parameter values $\alpha=-1.53^{+0.29}_{-0.19}$, $\overline{E}_{33}=1.55^{+0.26}_{-0.22}$  and  $\gamma=0.77\pm 0.24$. Here  $\alpha$ is   the  spectral index, $\gamma$ is the exponent of the Schechter luminosity function and $\overline{E}_{33}$  is the  mean FRB energy  in units of $10^{33} \, {\rm J}$ across   $2128 - 2848\; {\rm MHz}$ in the FRB rest frame.  
\end{abstract}


\section{Introduction}
Fast radio bursts (FRBs) are  milli-second duration,  highly  energetic ($\sim 10^{32}-10^{34}\,{\rm J}$) radio transients  \citep{lorimer07,keane11, thornton13}. Several radio telescopes including 
Parkes ( eg. \citealt{price18a}),  ASKAP (eg.  \citealt{bhandari20}), CHIME (eg. \citealt{amiri21}) and  UTMOST (eg. \citealt{gupta20}) have each detected a considerable number of FRBs.  Several  of the detected  FRBs are found to repeat (eg. \citealt{spitler14,amiri21}), however the non-repeating FRBs possibly form a  separate  population  \citep{Palaniswamy18,caleb18N,lu19} . 
Here we  only consider the non-repeating FRBs.  
The large dispersion measures (DMs), greater than the expected  Milky Way  contribution,  strongly suggests that FRBs are extragalactic events.  
Direct redshift estimates are available only for a few of the observed FRBs  which have been localised on  the sky (eg. \citealt{macquart20,heintz20}).  For most FRBs the  redshifts are  inferred from the observed DMs.

Several models have been proposed for the physical origin of  the FRB emission, unfortunately there is no  clear picture as yet. For example,  the recently  detected  FRB $200428$, which coincided  with an X-ray burst from  the  Galactic magnetar SGR $J1935+2154$  \citep{ridnaia21,li20,tavani21},  suggests active magnetars as a source for some of the FRBs  \citep{bochenek20, margalit20}. \citet{platts19} provides  a summary of the different FRB models.

The spectral index $\alpha$ of  the FRBs  is not very  well constrained at present.    \cite{macquart19} have  determined a mean value of $\alpha=-1.5^{+0.2}_{-0.3}$ for the sample of $23$ FRBs detected at ASKAP.   
\citet{houben19} have proposed a lower limit $\alpha>-1.2\pm-0.4$ considering the dearth of simultaneous detection of  FRB $121102$ at $1.4\,{\rm GHz}$ and $150\,{\rm MHz}$ respectively.   The energy and redshift distribution of the FRBs is also not well understood. \citet{james21} have modelled the FRB energy distribution using a simple power law . 
\citet{2021MNRAS.501..157Z} have used the FRBs detected at Parkes and ASKAP to constrain the exponent for the energy distribution to a value $-1.8$ . 
 
In an earlier work \citet{bera16} (hereafter Paper I) have  modelled the FRB  population and used this to make predictions for FRB detection at different telescopes. In a recent  work  \citet{bhattacharyya21} (hereafter Paper II) have used the two-dimensional  Kolmogorov-Smirnov (KS)  test to compare the FRBs observed at Parkes, ASKAP, CHIME and UTMOST with simulated  predictions for different FRB population models. It is shown there that the parameter range $\alpha>4$ and $\overline{E}_{33}>60$ is 
ruled out with $95\%$ confidence, here  $\overline{E}_{33}$ is the mean energy  of the FRBs population  in units of  $10^{33}\,{\rm J}$.   Paper II  also predicts that "CHIME is unlikely to
detect an FRB with extra-galactic dispersion measure $DM_{Ex}$ exceeding $3700\,{\rm pc\,cm^{-3}}$",  a prediction  which is borne out in the recently released CHIME catalogue of $492$ FRBs where the maximum value is $DM_{Ex}=3006.7\,{\rm pc\,cm^{-3}}$.  The modelling of the FRB population and simulations  of Paper II have also been used in the present work, and these are summarized  in the next section. In the present paper we have used a maximum likelihood analysis to estimate the parameters of the FRB population for which the predictions best match the observed FRB distribution. 

\section{Methodology}\label{sec:2}

For our analysis we use $82$ non-repeating FRBs detected by Parkes, ASKAP, CHIME and UTMOST which have each detected more than $10$ non-repeating FRBs.  
The frequency range, limiting fluence and number of FRBs for these four telescopes are summarized in the Table~\ref{tab:4_tel-ch6}. 
The recent  CHIME  data for  $492$  FRBs  \citep{rafiei21,amiri21} was released while this paper was being written, and we have not considered these here. 
Non-repeating  FRBs have also been detected at several other telescopes, however the number of events  at each of these telescopes is less than $10$  which is not adequate for the statistical analysis performed here.

\begin{figure}
\centering
\includegraphics[width=\columnwidth]{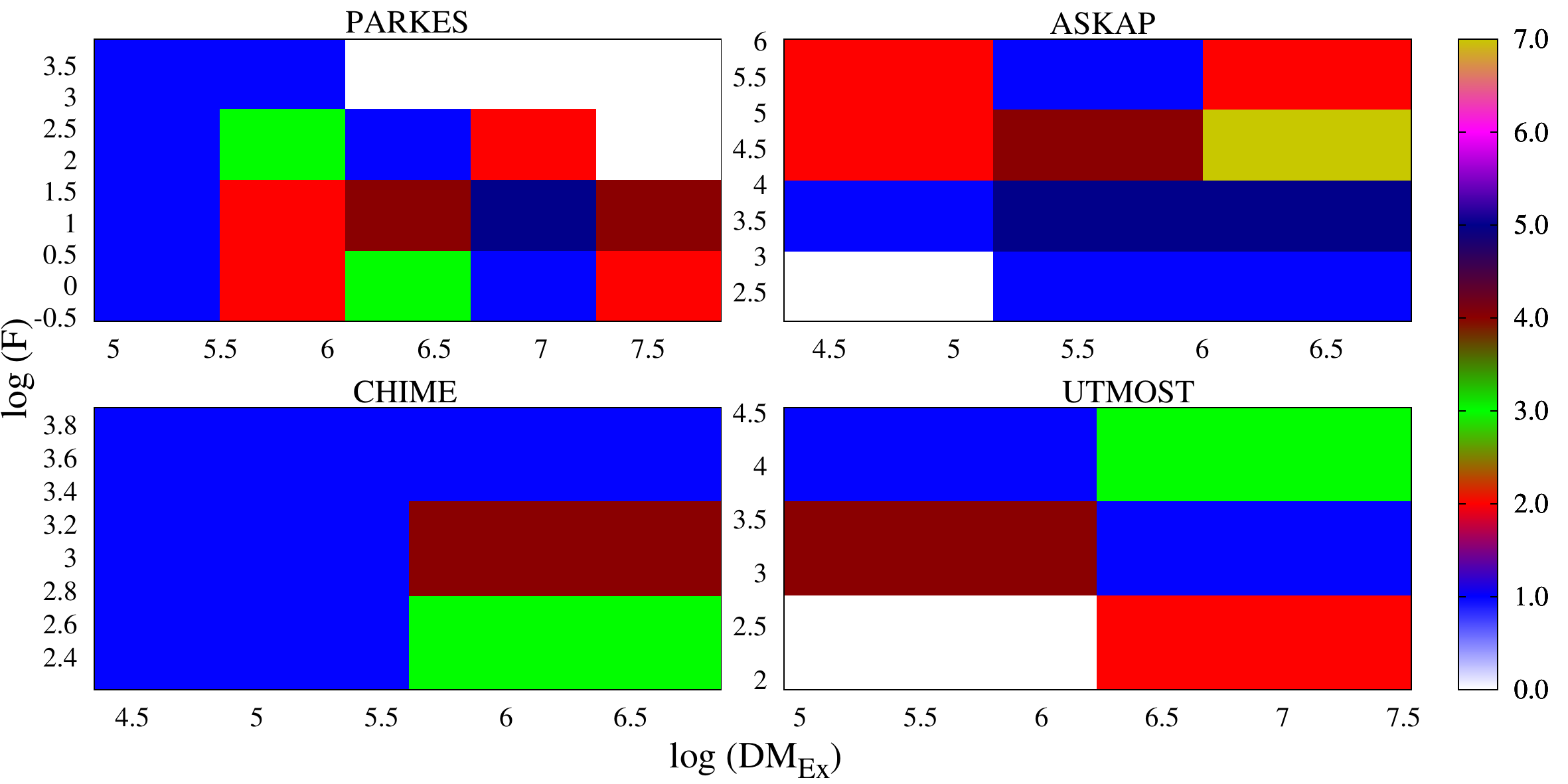}
\caption{Considering the FRB observations at Parkes, ASKAP, CHIME and UTMOST, this shows the number of observed FRBs $N_a$ at each grid point. The color scale shows the number of FRBs at each grid.}
\label{fig:grid_obs-ch6}
\end{figure}

Each observed FRB is characterized by its  dispersion measure $(DM)$, fluence $(F)$ and the pulse width $(w)$. Here we use the extragalactic component  $DM_{\rm Ex}=DM-DM_{\rm MW}$ where the Milky Way contribution is calculated for each FRB using the NE$2001$ model \citep{cordes03}.  For the present work  we have analysed  the observed distribution of $DM_{\rm Ex}$ and $F$ values. For each telescope, we have gridded the the observed  $DM_{\rm Ex}$ and $F$ range and calculated the number of observed FRBs $N_a$ at  each grid  point (Figure~\ref{fig:grid_obs-ch6}), the grid points here are labelled using   $a$. Given  the limited number of FRBs, it is necessary to use a very coarse grid for the present analysis.

\begin{table}
\caption{Considering the four telescopes Parkes; \citep{burke14,zhang20,zhang19,keane11,lorimer07,champion16, petroff19,thornton13,ravi15,petroff15,petroff17,keane16,bhandari18, ravi16,price18a,bhandari18a} ASKAP; \citep{bannister17,shannon18,macquart19,agarwal19,qiu19,bhandari19,bannister19,prochaska19,shannon19} CHIME; \citep{amiri19a} and UTMOST; \citep{caleb17,farah18,farah19,gupta20},  this shows the respective frequency range, limiting fluence $F_l$ and the number of FRBs detected there. Note that we have not included the 492 FRBs recently reported from CHIME. The value of $F_l$ depends on the threshold signal to noise ratio $(S/N)_{\rm th}$ which is different for the different FRB surveys.To keep the analysis simple, we consider a fixed $(S/N)_{\rm th}=10$.  Further,  ASKAP's sensitivity also  differs  based on the particular survey in which the FRB was detected. For simplicity, we have used a fixed value of $F_l$ for ASKAP.}
    \centering
    \begin{tabular}{cccc}
    \hline
    Telescope & Frequency range & $F_l$ & Number of \\
    Name & $({\rm MHz})$ & $({\rm Jy\,ms})$ & non-repeating FRBs \\
    \hline
    Parkes & $1157-1546$ & $0.5$ & $29$ \\
    ASKAP & $1129-1465$ & $4.1$ & $31$ \\
    CHIME & $400-800$ & $0.64$ & $11$ \\
    UTMOST & $827-859$ & $3.25$ & $11$ \\
    \hline
    \end{tabular}
    \label{tab:4_tel-ch6}
\end{table}
We now briefly discuss our model for the FRB population. 
The model for the FRB population is presented in Paper I, and the reader is referred there for details. The intrinsic properties of each FRB event are  characterised by three quantities $\alpha$,  $E_{33}$ and $w_i$. Here the energy of the FRB pulse at any frequency $\nu$ is assumed to be proportional to $\nu^\alpha$ where $\alpha$  is the spectral index. For the present analysis we have assumed that all the FRBs have the same value of the spectral index $\alpha$. Here $E_{33}$ is the energy  of the  FRB (in units of $10^{33} \, {\rm J}$) emitted in the frequency interval $2128\,{\rm MHz}$ to $2848\,{\rm MHz}$ at the rest frame of the source. $w_i$ here is  the  intrinsic pulse width  of the FRB. 
Our earlier work (Paper II) shows that the results do not change much if we vary $w_i$ in the range $0.1 \, {\rm ms}$ to $2.0 \, {\rm ms}$, and here we have used a fixed value $w_i=1\,{\rm ms}$ for  the entire analysis.  In addition, each FRB also has a redshift $z$ and an angular position $\mathbf{\theta}$ on the sky.

The FRB energy distribution is currently unknown. It is reasonable to assume that FRBs have a characteristic energy $\overline{E}_{33}$, and the distribution falls of rapidly beyond $\overline{E}_{33}$.   For the analysis presented here we have assumed a Schechter luminosity function \citep{schechter76} for the energy distribution expressed as:
\begin{equation}
    n(E_{33},z)=%
    \begin{cases}
    \frac{n_0(z)}{\overline{E_{33}}}\;\exp\left(-\frac{E_{33}}{\overline{E}_{33}}\right) & {\rm for}\;\gamma=0 \\
    \frac{n_0(z)}{\overline{E}_{33}}\left(\frac{\gamma}{\Gamma(1+\gamma)}\right)\left(\frac{\gamma E_{33}}{\overline{E}_{33}}\right)^{\gamma} \exp\left(-\frac{\gamma E_{33}}{\overline{E}_{33}}\right) & {\rm for}\;\gamma>0
    \end{cases}
    \label{eq:Sch_func-ch6}
\end{equation}
where $\overline{E}_{33}$ is the mean energy of the population, $\gamma$ is the exponent and $\Gamma(n)$ is the Gamma function. Here $n(E_{33},z)$ is the FRB event rate per unit comoving volume per energy interval $d E_{33}$  and $n_0(z)$ is  the FRB event rate per unit comoving volume. 
In our analysis, we have considered two possibilities for the event rate evolution with redshift, namely (a.) CER -  constant event rate  where $n_0(z)$ is independent of $z$ over the redshift range of our interest and (b.) SFR -  where $n_0(z)\propto(0.015(1+z)^{2.7})/(1+((1+z)/2.9)^{5.6})$ traces the star formation rate over cosmic time \citep{madau14}.

Considering an FRB located at redshift $z$, the observed pulse width $w=\sqrt{w_{\rm cos}^2+w_{\rm DM}^2+w_{\rm sc}^2}$ mainly has three contributions, namely (a) the cosmic expansion $w_{\rm cos}=w_i\,(1+z)$ - the intrinsic pulse width $w_i$, expanded by the cosmological expansion, (b) the dispersion broadening $w_{\rm DM}=(8.3\times10^6\,{\rm DM}\,\Delta\nu_c)/\nu_0^3$ - the residual dispersion broadening after the incoherent dedispersion of the FRB signal where $\nu_0$ is the observational frequency and $\Delta\nu_c$ is the channel width of the telescope, and (c) the scatter broadening $w_{\rm sc}$ - this is not well understood to date. Here  $w$ is expressed in ${\rm ms}$, and both $\nu_0$ and $\Delta\nu_c$ are expressed in $\rm MHz$. 
In this work we have considered three scattering models, namely (a) Sc-I - based on the empirical fit of a large number of Galactic pulsar data provided  by \cite{bhat04} and we have extrapolated this for the IGM, (b) Sc-II - this is a pure analytical model proposed by \cite{macquart13} considering the turbulent IGM, and (c) No-Sc - where there is no scattering and thus $w_{\rm sc}=0$. Our earlier work (Figure~1 of Paper-I) shows that for both Sc-I and -II,  scattering dominates the total pulse width at redshifts $z>0.5$. Further, 
the pulse width  increases very sharply with increasing redshift for Sc-I relative to Sc-II. 
The pulse width increases very slowly with redshift for No-Sc.  

The contribution  $DM_{\rm Host}$ from the host galaxy  is an unknown factor that enters FRB observations  for most of the unlocalized FRBs. The value of $DM_{\rm Host}$ is expected to vary from FRB to FRB depending on the host galaxy and  the location of the FRB within it. 
However, the FRB detections suggest that $DM_{\rm Host}$ may not exceed the value $100\,{\rm pc\,cm}^{-3}$ \citep{macquart20}. In addition to this, the observed DM will  have another contribution $(DM_{\rm Halo} \sim 50-80\,{\rm pc\,cm}^{-3})$ from the Galactic halo  \citep{prochaska19a}. Here we have  absorbed  the $DM_{\rm Halo}$ contribution in  $DM_{\rm Host}$.
In this work we consider two scenarios for $DM_{\rm Host}$ namely (a) DM120 - where all FRBs have fixed $DM_{\rm Host}=120\,{\rm pc\,cm}^{-3}$, and 
(b.) DMRand - where $DM_{\rm Host}$ values are randomly  drawn from a Gaussian distribution with mean $\overline{DM}_{\rm Host}= 120\,{\rm pc\,cm}^{-3}$ and root mean square value $\Delta {DM}_{\rm Host}= 15 \,{\rm pc\,cm}^{-3}$. The $DM_{\rm Host}$ distribution is truncated at $0$.

In  summary, our model for the FRB population has three parameters $\alpha$, $\overline{E}$ and $\gamma$. Further, we have two models for the event rate distribution namely SFR and CER, three models for the scattering namely Sc-I, Sc-II and No-Sc, and two models for $DM_{\rm Host}$  namely DM120 and DMRand.

\begin{figure}
\centering
\includegraphics[width=\columnwidth]{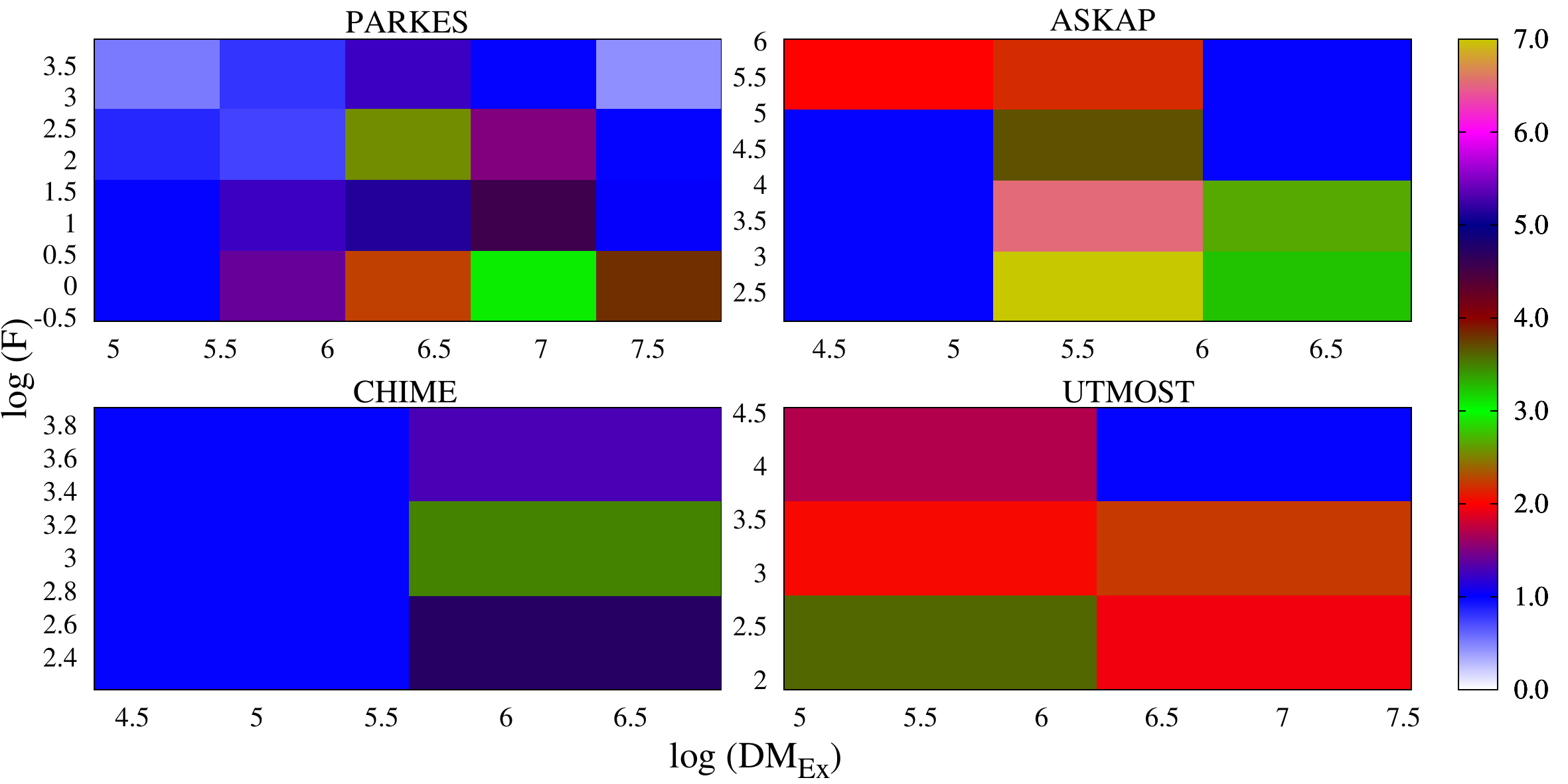}
\caption{Considering Parkes, ASKAP, CHIME and UTMOST, this shows $\mu_a$ the mean number of  FRBs predicted  at each grid point for a model with  $\alpha=-1.53$, $\overline{E}_{33}=1.55\,{\rm J}$ and $\gamma=0.77$ with Sc-II, CER and DM120. The color scale shows the number of FRBs at each grid.}
\label{fig:grid_pred-ch6}
\end{figure}

We now briefly discuss how we use simulations   to calculate $\mu_a$  (Figure \ref{fig:grid_pred-ch6})  which is the model prediction for the mean number of FRBs expected to be detected at each grid point for any given telescope. These simulations are discussed  in Paper II, and the interested readers are requested to refer to it for more details. We consider a comoving volume which extends up to $z_{\rm max}=5$, which is considerably higher than the highest redshift inferred for any of the FRB events included in our analysis. The angular extent of this comoving volume is equal to the full width half maxima (FWHM) of the primary beam of the telescope and this varies from telescope to telescope.
We populate this comoving volume with $10^6$ randomly located FRBs whose mean comoving number density follow $n_0(z)$. This provides the comoving distance $r$  and angular position $\mathbf{\theta}$ for each simulated FRB. 
We consider the $\Lambda CDM$ cosmology \citep{planck20} to calculate  $z$ from $r$. The energy $E_{33}$ of each FRB  is randomly drawn from the distribution in  eq.~(\ref{eq:Sch_func-ch6}). We calculate $DM_{\rm Ex}$, $F$ and $w$ for each simulated FRB. 
The telescope can only detect an FRB if $F\times \sqrt{1{\rm ms}/w}\geq F_l$  where $F_l$ is the limiting fluence of the telescope which depends on the threshold signal to noise ratio $(S/N)_{\rm th}$.  Considering $(S/N)_{\rm th}=10$,   Table~\ref{tab:4_tel-ch6} lists the value of $F_l$ for the four telescopes considered here.  We determine the fraction of observable events corresponding to  each ($DM_{\rm Ex}, F$) grid point  and multiply this with the total number of FRBs actually observed by the telescope to calculate $\mu_a$ which is the mean number of FRB's expected in each grid point for the particular model under consideration. Figure ~\ref{fig:grid_pred-ch6} shows $\mu_a$ predicted for a particular model $\alpha=-1.53$, $\overline{E}_{33}=1.55$ and $\gamma=0.77$ with Sc-II, CER and DM120. We have compared the simulated predictions ($\mu_a$) with the actual observation ($N_a$) to identify preferred models for the FRB population.

We have used a Bayesian inference framework to constrain the parameters $\alpha$, $\overline{E}_{33}$ and $\gamma$ for all FRB models discussed here. The likelihood is estimated using the Poisson statistic. Considering the $a^{\rm th}$ grid the probability of getting $N_a$ for a given $\mu_a$ is 
expressed as 
\begin{equation}
    P_a=\frac{\mu_a^{N_a}\exp{(-\mu_a)}}{N_a!}
    \label{eq:la-ch6}
\end{equation}
where we assume that the likelihood $l_a$ is proportional to the probability, \textit{i.e.} $l_a\propto P_a$. In this analysis we choose the proportionality constant to be $1$.  
Considering all the grid points, the logarithmic value of the total likelihood is given by
\begin{equation}
    \log(\mathcal{L})=\sum_{a} \log(l_a)=\sum_{a}\left[N_a\log(\mu_a)-\mu_a-\log(N_a!)\right]
    \label{eq:logL-ch6}
\end{equation}
where the summation is taken over all the grid points. 
Our aim here is to probe efficiently the maximum likely region of the parameter space using a Markov-Chain-Monte-Carlo (MCMC) algorithm. It has been demonstrated by \citet{autcha} that in case of a Poissonian likelihood this can done efficiently by estimating the loss function:
\begin{equation}
   L(\mu_a,N_a) = -\sum_a\left[\mu_a-N_a\right]^{T}.\left[\mu_a-N_a\right]
   \label{eq:logL-ch7}
\end{equation}
 at each step of the random walker in the parameter space. The evaluation of this loss function at each step of the random walk can efficiently guide the random walker towards the  most likely region of the parameter space.  
For a detailed discussion on this, the interested readers are referred to \citet{autcha}.
In this work we have used the \texttt{Python}-based package \texttt{Emcee}\footnote{Publicly available at: \href{https://pypi.org/project/emcee/}{https://pypi.org/project/emcee/}} \citep{Foreman_Mackey_2013}-- an Affine invariant MCMC ensemble sampler \citep{Goodman_2010} to perform the exploration of the parameter space.
For a specific FRB model and the given set of model parameters $(\alpha, \overline{E}_{33}, \gamma)$ we first predict $\mu_a$ and then using $N_a$ and $\mu_a$ we estimate the loss function $L(\mu_a,N_a)$ that eventually leads to estimation of the $\log(\mathcal{L})$ of the region around the maximum likelihood. For this analysis, we have used a standard MCMC chain of $20,000$ samples with $40$ random walkers. We discarded the initial $10$ per cent of the $20,000$ samples as the burn-in steps. We have varied our parameter search between the range $\alpha\rightarrow (-30.0,+30.0), \overline{E}_{33}\rightarrow (0.01,20.00), \gamma\rightarrow (0.0,6.0)$, and assumed a uniform prior while doing the estimations.

\section{Results}\label{sec:3}
Table~\ref{tab:likelihood-ch6}  shows the best fit  parameter values for all the models which we have considered here, the maximum values of log-likelihood $\log(\mathcal{L})$ are also shown alongside for reference. Considering the models with Sc-I, comparing the $\log(\mathcal{L})$ values we see that the models with CER are preferred over the SFR models.  
The same feature is also seen  for the other scattering models. Guided by this, we exclude the SFR models and  focus entirely on the CER models  for  the subsequent discussion. 
Comparing the different scattering models next, we find that Sc-II is preferred over both Sc-I and No-Sc. Restricting our attention to Sc-II with CER, we find that DM120 and DMRand have  comparable $\log(\mathcal{L})$ values. 
Although the $\log(\mathcal{L})$ value is slightly larger for DM120, the difference between DM120 and DMRand is very small in comparison to the difference with all the other models.   

\begin{figure}
\centering
\includegraphics[width=\columnwidth]{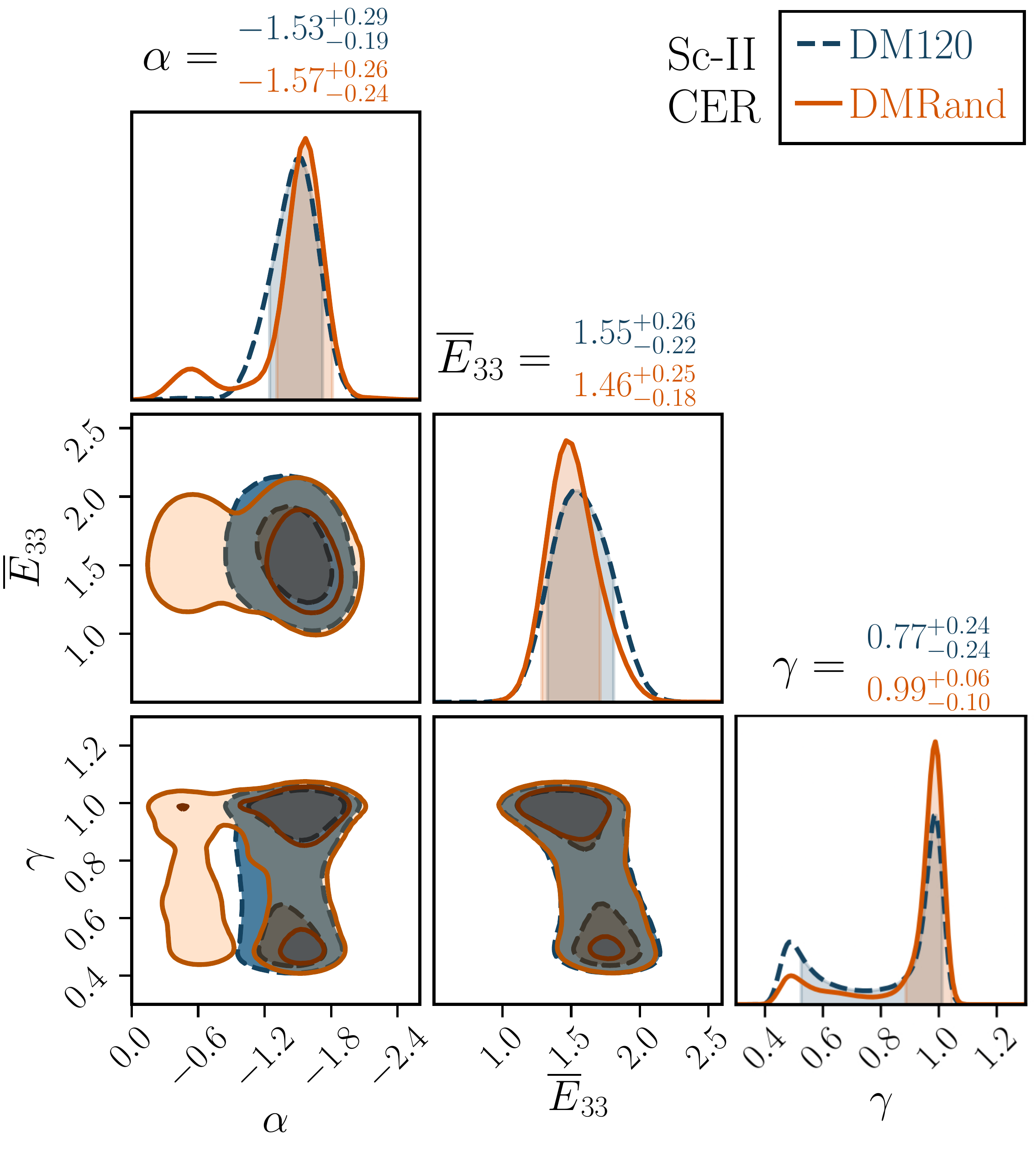}
\caption{Considering the FRB observations at Parkes, ASKAP, CHIME and UTMOST, this shows the maximum likelihood contours for the parameters $\alpha, \overline{E}_{33}$ and $\gamma$ considering models with  Sc-II and CER. The 2D plot show both $68\%$ and  $95\%$ confidence intervals. Only the $ 68\%$  confidence interval is shown in the 1D marginalized plots. The best fit values of $\alpha$, $\overline{E}_{33}$ and $\gamma$ are shown in the figure with blue color for DM120 and orange color for DMRand respectively.}
\label{fig:joint_cer_sc2-ch6}
\end{figure}

Considering Sc-II with CER, Figure~\ref{fig:joint_cer_sc2-ch6} shows a corner plot of  the parameters  $\alpha$, $\overline{E}_{33}$ and $\gamma$  for both DM120 and DMRand. We first consider the three plots which show the joint distribution of pairs of these  parameters.  A visual inspection shows that in all cases the $68 \%$ and  $95 \%$   confidence intervals  for DM120 largely overlap with those  for DMRand. Further, the orientation of the confidence intervals indicates that the constraints on the three parameters $\alpha$, $\overline{E}_{33}$ and $\gamma$ are largely uncorrelated. 
For DMRand for ($\alpha,\gamma$)  we notice that the $68 \%$ confidence interval is divided into two disconnected regions both of which are enclosed within the $95 \%$   confidence interval. Here also one of the region dominates whereas  the other has a very low probability associated with it.
Considering the two dimensional parameter plots,  we see that 
in all cases  our analysis imposes tight constraints on the joint distribution of the parameters  $\alpha$, $\overline{E}_{33}$ and $\gamma$.
We now consider the marginalised one dimensional plots which correspond to the best fit parameter values and the $68 \%$ confidence intervals reported in Table~\ref{tab:likelihood-ch6}. 
We find that $\alpha$ is well constrained by our analysis with $\alpha=-1.53^{+0.29}_{-0.19}$  and $-1.57^{+0.26}_{-0.24}$ for DM120 and DMRand respectively. Similarly, we find that $\overline{E}_{33}$ is  well constrained with  $\overline{E}_{33}=1.55^{+0.26}_{-0.22}$ and $1.46^{+0.25}_{-0.18}$ for DM120 and DMRand respectively.
We also find that $\gamma$ is well constrained with $\gamma=0.77\pm 0.24$ and $0.988^{+0.058}_{-0.102}$ for DM120 and DMRand respectively.
For $\alpha$, $\overline{E}_{33}$ and $\gamma$, the probability distributions are dominated by a single peak. Further, we see that the probability distributions for DM120 and DMRand have considerable overlap which indicates that the $\alpha$, $\overline{E}_{33}$ and $\gamma$ values estimated for these two models are consistent with one another.

\section{Conclusions}
In this analysis we have considered two distinct scenarios, one where the comoving FRB event rate evolves with redshift following the star formation rate  and another where it is constant independent of redshift. Our analysis shows that the models with a constant event rate are preferred over the models where the event rate follows the star formation rate. The pulse broadening due to scattering in the IGM is not clearly understood at present. Here we have considered three possibilities. 
In the first model,  based on \cite{bhat04} which provides an empirical fit of a large number of Galactic pulsar data, the scattering pulse width increases very sharply with increasing redshift. 
The  second one is  a purely  theoretical model proposed by \cite{macquart13} considering the turbulent IGM. In this model the scattering pulse has a modest  increase with increasing redshift. 
The third model assumes no scattering.  Our analysis shows that the second model where there is  a modest increase in pulse width with increasing redshift is preferred to the model where this increase is much steeper or the model where there is no scattering. 
The contribution to the dispersion measure from the host galaxies $DM_{\rm Host}$ is another quantity which affects FRB detections.  
Here we consider two models, one where all the   FRBs have fixed $DM_{\rm Host}=120\,{\rm pc\,cm}^{-3}$, and another where the $DM_{\rm Host}$ values are randomly  drawn from a truncated Gaussian distribution with mean $\overline{DM}_{\rm Host}= 120\,{\rm pc\,cm}^{-3}$ and root mean square value $\Delta {DM}_{\rm Host}= 15 \,{\rm pc\,cm}^{-3}$.  
Our analysis shows that the model with a fixed $DM_{\rm Host}$ is slightly preferred, however we find comparable results from both the models. Considering the preferred combination of the event rate distribution and  scattering,  we have $(\alpha, \overline{E}_{33}, \gamma)= (-1.53^{+0.29}_{-0.19},1.55^{+0.26}_{-0.22},0.77\pm 0.24)$ and $(-1.57^{+0.26}_{-0.24},1.46^{+0.25}_{-0.18},0.988^{+0.058}_{-0.102})$ for DM120 and DMRand respectively. 
The Schechter luminosity function approaches a Dirac-delta function as the value of $\gamma$ is increased. The relatively low value of $\gamma$ indicates that there is a considerable spread in the values of the FRB energies. 
We have assumed $w_i= 1 \, {\rm ms}$ for the entire analysis presented here. The analysis was repeated for  $w_i= 0.5 \, {\rm ms}$ for which the change in the best fit values  was found to be less than $10\%$.

An earlier work (Paper II) had used the KS   test to rule out  the parameter range $\alpha>4$ and $\overline{E}_{33}>60$   with $95\%$ confidence. We have checked  that the best fit parameter values obtained here are well within the allowed parameter range identified in Paper II.

The results of this paper are expected to provide inputs for any physical model for the nature of the FRB sources and their cosmological distribution. It also throws some light on the effect of scattering in the IGM. We plan to carry out a similar analysis using the recently released catalogue of $492$ FRBs detected at CHIME.  We expect tighter constraints as more FRBs get included in the analysis. 

\renewcommand{\arraystretch}{1.7}
\begin{table*}
    \caption{Considering the FRB detected  at Parkes, ASKAP, CHIME and UTMOST, this shows the best fit  values along with $68\%$ confidence intervals for  $\alpha$, $\overline{E}_{33}$  and $\gamma$ for the different models considered here. The maximum values of log-likelihood $[\log(\mathcal{L})]_{\rm max}$ are also shown for reference. Results are shown for two different models  for $DM_{\rm Host}$  namely DM120 and DMRand.}
    \centering
    \begin{tabular}{cccccccccc}
    \hline
     & & & DM120 & & & & DMRand & & \\
    
    \hline
     & FRB & \multicolumn{3}{c}{Parameter Values} & Value of & \multicolumn{3}{c}{Parameter Values} & Value of \\
     & Rate & $\alpha$ & $\overline{E}_{33}$ & $\gamma$ & $[\log(\mathcal{L})]_{\rm max}$ & $\alpha$ & $\overline{E}_{33}$ & $\gamma$ & $[\log(\mathcal{L})]_{\rm max}$ \\
    \hline
    Sc-I&CER & $-0.50^{+0.25}_{-0.27}$ & $2.18^{+0.29}_{-0.29}$ &$0.96^{+0.04}_{-0.40}$ & $-129.64$ &  $-0.58^{+0.19}_{-0.15}$ & $2.19^{+0.27}_{-0.25}$ & $0.89^{+0.12}_{-0.52}$ & $-133.08$ \\
    &SFR & $-26.0^{+1.8}_{-1.6}$ & $4.82^{+0.98}_{-0.76}$ & $0.31^{+0.38}_{-0.11}$ & $-235.57$ & $-25.8^{+1.7}_{-1.5}$ & $5.01^{+0.86}_{-0.84}$ & $0.26^{+0.40}_{-0.05}$ & $-235.39$ \\
    \hline
Sc-II&CER & $-1.53^{+0.29}_{-0.19}$ & $1.55^{+0.26}_{-0.22}$
    &$0.77^{+0.24}_{-0.24}$ &$-78.39$ & $-1.57^{+0.26}_{-0.24}$ & $1.46^{+0.25}_{-0.18}$ & $0.99^{+0.06}_{-0.10}$& $-79.95$\\
    &SFR & $-27.5^{+1.5}_{-1.7}$ & $3.96^{+0.76}_{-0.70}$ & $0.37^{+0.41}_{-0.16}$ & $-233.31$ & $-27.5^{+1.5}_{-1.6}$ & $4.03^{+0.63}_{-0.68}$ & $0.29^{+0.43}_{-0.05}$ & $-233.18$ \\
    \hline
No-Sc&CER & $-2.26^{+0.19}_{-0.20}$ & $1.09^{+0.19}_{-0.24}$ & $0.56^{+0.34}_{-0.12}$ & $-94.82$ &  $-2.25^{+0.12}_{-0.15}$ & $1.09^{+0.07}_{-0.08}$ & $0.47^{+0.33}_{-0.02}$ & $-95.29$ \\
    &SFR & $-30.0^{+1.1}_{-0.0}$ & $2.81^{+0.61}_{-0.56}$ & $0.40^{+0.49}_{-0.16}$ & $-229.16$ & $-30.0^{+1.2}_{-0.0}$ & $2.80^{+0.58}_{-0.56}$ & $0.40^{+0.50}_{-0.17}$ & $-229.34$  \\
    \hline    
    \end{tabular}
    \label{tab:likelihood-ch6}
\end{table*}

\section*{Acknowledgement}
We acknowledge the Supercomputing facility ‘PARAM-Shakti’ at  IIT Kharagpur established under the  National Supercomputing Mission (NSM), Government of India and supported by Centre for Development of Advanced Computing (CDAC), Pune. Some part of the statistical analysis for this work was done using the computing resources available to the Cosmology with Statistical Inference (CSI) research group at IIT Indore.

\section*{Data availability}
The data and codes underlying this article will be shared on reasonable request to the corresponding author.

\bibliographystyle{mnras}
\bibliography{reference}

\begin{thebibliography}{}
\makeatletter
\relax
\def\mn@urlcharsother{\let\do\@makeother \do\$\do\&\do\#\do\^\do\_\do\%\do\~}
\def\mn@doi{\begingroup\mn@urlcharsother \@ifnextchar [ {\mn@doi@}
  {\mn@doi@[]}}
\def\mn@doi@[#1]#2{\def\@tempa{#1}\ifx\@tempa\@empty \href
  {http://dx.doi.org/#2} {doi:#2}\else \href {http://dx.doi.org/#2} {#1}\fi
  \endgroup}
\def\mn@eprint#1#2{\mn@eprint@#1:#2::\@nil}
\def\mn@eprint@arXiv#1{\href {http://arxiv.org/abs/#1} {{\tt arXiv:#1}}}
\def\mn@eprint@dblp#1{\href {http://dblp.uni-trier.de/rec/bibtex/#1.xml}
  {dblp:#1}}
\def\mn@eprint@#1:#2:#3:#4\@nil{\def\@tempa {#1}\def\@tempb {#2}\def\@tempc
  {#3}\ifx \@tempc \@empty \let \@tempc \@tempb \let \@tempb \@tempa \fi \ifx
  \@tempb \@empty \def\@tempb {arXiv}\fi \@ifundefined
  {mn@eprint@\@tempb}{\@tempb:\@tempc}{\expandafter \expandafter \csname
  mn@eprint@\@tempb\endcsname \expandafter{\@tempc}}}

\bibitem[\protect\citeauthoryear{Agarwal, Lorimer  et~al.}{Agarwal
  et~al.}{2019}]{agarwal19}
Agarwal D.,  Lorimer D.~R.,   et~al., 2019, MNRAS, 490, 1

\bibitem[\protect\citeauthoryear{Aghanim, Akrami  et~al.}{Aghanim
  et~al.}{2020}]{planck20}
Aghanim N.,  Akrami Y.,   et~al., 2020, A\&A, 641, A12

\bibitem[\protect\citeauthoryear{Amiri, Bandura  et~al.}{Amiri
  et~al.}{2019}]{amiri19a}
Amiri M.,  Bandura K.,   et~al., 2019, Nature, 566, 230

\bibitem[\protect\citeauthoryear{Amiri, Andersen  et~al.}{Amiri
  et~al.}{2021}]{amiri21}
Amiri M.,  Andersen B.~C.,   et~al., 2021, arXiv preprint, 2106.04352

\bibitem[\protect\citeauthoryear{Autcha}{Autcha}{2014}]{autcha}
Autcha A.,  2014, Science \&; Technology Asia, 19, 14

\bibitem[\protect\citeauthoryear{Bannister, Shannon  et~al.}{Bannister
  et~al.}{2017}]{bannister17}
Bannister K.~W.,  Shannon R.~M.,   et~al., 2017, ApJ Letters, 841, L12

\bibitem[\protect\citeauthoryear{Bannister, Deller  et~al.}{Bannister
  et~al.}{2019}]{bannister19}
Bannister K.~W.,  Deller A.~T.,   et~al., 2019, Science, 365, 565

\bibitem[\protect\citeauthoryear{Bera, Bhattacharyya  et~al.}{Bera
  et~al.}{2016}]{bera16}
Bera A.,  Bhattacharyya S.,   et~al., 2016, MNRAS, 457, 2530

\bibitem[\protect\citeauthoryear{Bhandari, Keane  et~al.}{Bhandari
  et~al.}{2018a}]{bhandari18}
Bhandari S.,  Keane E.~F.,   et~al., 2018a, MNRAS, 475, 1427

\bibitem[\protect\citeauthoryear{Bhandari, Caleb  et~al.}{Bhandari
  et~al.}{2018b}]{bhandari18a}
Bhandari S.,  Caleb M.,   et~al., 2018b, The Astronomer's Telegram, 12060, 1

\bibitem[\protect\citeauthoryear{Bhandari, Bannister  et~al.}{Bhandari
  et~al.}{2019}]{bhandari19}
Bhandari S.,  Bannister K.~W.,   et~al., 2019, MNRAS, 486, 70

\bibitem[\protect\citeauthoryear{Bhandari, Sadler  et~al.}{Bhandari
  et~al.}{2020}]{bhandari20}
Bhandari S.,  Sadler E.~M.,   et~al., 2020, ApJ Letters, 895, L37

\bibitem[\protect\citeauthoryear{Bhat, Cordes  et~al.}{Bhat
  et~al.}{2004}]{bhat04}
Bhat N. D.~R.,  Cordes J.~M.,   et~al., 2004, ApJ, 605, 759

\bibitem[\protect\citeauthoryear{Bhattacharyya \& Bharadwaj}{Bhattacharyya \&
  Bharadwaj}{2021}]{bhattacharyya21}
Bhattacharyya S.,  Bharadwaj S.,  2021, MNRAS, 502, 904

\bibitem[\protect\citeauthoryear{Bochenek, Ravi  et~al.}{Bochenek
  et~al.}{2020}]{bochenek20}
Bochenek C.~D.,  Ravi V.,   et~al., 2020, Nature, 587, 59

\bibitem[\protect\citeauthoryear{Burke-Spolaor \& Bannister}{Burke-Spolaor \&
  Bannister}{2014}]{burke14}
Burke-Spolaor S.,  Bannister K.~W.,  2014, ApJ, 792, 19

\bibitem[\protect\citeauthoryear{Caleb, Flynn  et~al.}{Caleb
  et~al.}{2017}]{caleb17}
Caleb M.,  Flynn C.,   et~al., 2017, MNRAS, 468, 3746

\bibitem[\protect\citeauthoryear{Caleb, Spitler  \& Stappers}{Caleb
  et~al.}{2018}]{caleb18N}
Caleb M.,  Spitler L.~G.,   Stappers B.~W.,  2018, Nature Astronomy, 2, 839

\bibitem[\protect\citeauthoryear{Champion, Petroff  et~al.}{Champion
  et~al.}{2016}]{champion16}
Champion D.~J.,  Petroff E.,   et~al., 2016, MNRAS: Letters, 460, L30

\bibitem[\protect\citeauthoryear{Cordes \& Lazio}{Cordes \&
  Lazio}{2003}]{cordes03}
Cordes J.~M.,  Lazio T. J.~W.,  2003, arXiv preprint, astro-ph/0207156

\bibitem[\protect\citeauthoryear{Farah, Flynn  et~al.}{Farah
  et~al.}{2018}]{farah18}
Farah W.,  Flynn C.,   et~al., 2018, MNRAS, 478, 1209

\bibitem[\protect\citeauthoryear{Farah, Flynn  et~al.}{Farah
  et~al.}{2019}]{farah19}
Farah W.,  Flynn C.,   et~al., 2019, MNRAS, 488, 2989

\bibitem[\protect\citeauthoryear{Foreman-Mackey, Hogg  et~al.}{Foreman-Mackey
  et~al.}{2013}]{Foreman_Mackey_2013}
Foreman-Mackey D.,  Hogg D.~W.,   et~al., 2013, PASP, 125, 306–312

\bibitem[\protect\citeauthoryear{Goodman \& Weare}{Goodman \&
  Weare}{2010}]{Goodman_2010}
Goodman J.,  Weare J.,  2010, Comm. App. Math. Com. Sc., 5, 65

\bibitem[\protect\citeauthoryear{Gupta, Bailes  et~al.}{Gupta
  et~al.}{2020}]{gupta20}
Gupta V.,  Bailes M.,   et~al., 2020, The Astronomer's Telegram, 13788, 1

\bibitem[\protect\citeauthoryear{Heintz, Prochaska  et~al.}{Heintz
  et~al.}{2020}]{heintz20}
Heintz K.~E.,  Prochaska J.~X.,   et~al., 2020, ApJ, 903, 152

\bibitem[\protect\citeauthoryear{Houben, Spitler  et~al.}{Houben
  et~al.}{2019}]{houben19}
Houben L. J.~M.,  Spitler L.~G.,   et~al., 2019, A \& A, 623, A42

\bibitem[\protect\citeauthoryear{James, Prochaska  et~al.}{James
  et~al.}{2021}]{james21}
James C.~W.,  Prochaska J.~X.,   et~al., 2021, arXiv preprint, 2101.08005

\bibitem[\protect\citeauthoryear{Keane, Kramer  et~al.}{Keane
  et~al.}{2011}]{keane11}
Keane E.~F.,  Kramer M.,   et~al., 2011, MNRAS, 415, 3065

\bibitem[\protect\citeauthoryear{Keane, Johnston  et~al.}{Keane
  et~al.}{2016}]{keane16}
Keane E.~F.,  Johnston S.,   et~al., 2016, Nature, 530, 453

\bibitem[\protect\citeauthoryear{Li et~al.,}{Li et~al.}{2020}]{li20}
Li C.~K.,  et~al., 2020, arXiv preprint, 2005.11071

\bibitem[\protect\citeauthoryear{Lorimer, Bailes  et~al.}{Lorimer
  et~al.}{2007}]{lorimer07}
Lorimer D.~R.,  Bailes M.,   et~al., 2007, Science, 318, 777

\bibitem[\protect\citeauthoryear{Lu \& Piro}{Lu \& Piro}{2019}]{lu19}
Lu W.,  Piro A.~L.,  2019, ApJ, 883, 40

\bibitem[\protect\citeauthoryear{Macquart \& Koay}{Macquart \&
  Koay}{2013}]{macquart13}
Macquart J.~P.,  Koay J.~Y.,  2013, ApJ, 776, 125

\bibitem[\protect\citeauthoryear{Macquart, Shannon  et~al.}{Macquart
  et~al.}{2019}]{macquart19}
Macquart J.~P.,  Shannon R.~M.,   et~al., 2019, ApJ Letters, 872, L19

\bibitem[\protect\citeauthoryear{Macquart et~al.,}{Macquart
  et~al.}{2020}]{macquart20}
Macquart J.~P.,  et~al., 2020, Nature, 581, 391

\bibitem[\protect\citeauthoryear{Madau \& Dickinson}{Madau \&
  Dickinson}{2014}]{madau14}
Madau P.,  Dickinson M.,  2014, Annual Review of A \& A, 52, 415

\bibitem[\protect\citeauthoryear{Margalit, Beniamini  et~al.}{Margalit
  et~al.}{2020}]{margalit20}
Margalit B.,  Beniamini P.,   et~al., 2020, ApJ Letters, 899, L27

\bibitem[\protect\citeauthoryear{Palaniswamy, Li  \& Zhang}{Palaniswamy
  et~al.}{2018}]{Palaniswamy18}
Palaniswamy D.,  Li Y.,   Zhang B.,  2018, ApJ Letters, 854, L12

\bibitem[\protect\citeauthoryear{Petroff, Bailes  et~al.}{Petroff
  et~al.}{2015}]{petroff15}
Petroff E.,  Bailes M.,   et~al., 2015, MNRAS, 447, 246

\bibitem[\protect\citeauthoryear{Petroff, Burke-Spolaor  et~al.}{Petroff
  et~al.}{2017}]{petroff17}
Petroff E.,  Burke-Spolaor S.,   et~al., 2017, MNRAS, 469, 4465

\bibitem[\protect\citeauthoryear{Petroff, Oostrum  et~al.}{Petroff
  et~al.}{2019}]{petroff19}
Petroff E.,  Oostrum L.~C.,   et~al., 2019, MNRAS, 482, 3109

\bibitem[\protect\citeauthoryear{Platts, Weltman  et~al.}{Platts
  et~al.}{2019}]{platts19}
Platts E.,  Weltman A.,   et~al., 2019, Physics Reports, 821, 1

\bibitem[\protect\citeauthoryear{Price et~al.,}{Price et~al.}{2018}]{price18a}
Price D.~C.,  et~al., 2018, The Astronomer's Telegram, 11376, 1

\bibitem[\protect\citeauthoryear{Prochaska \& Zheng}{Prochaska \&
  Zheng}{2019}]{prochaska19a}
Prochaska J.~X.,  Zheng Y.,  2019, MNRAS, 485, 648

\bibitem[\protect\citeauthoryear{Prochaska et~al.,}{Prochaska
  et~al.}{2019}]{prochaska19}
Prochaska J.~X.,  et~al., 2019, Science, 366, 231

\bibitem[\protect\citeauthoryear{Qiu, Bannister  et~al.}{Qiu
  et~al.}{2019}]{qiu19}
Qiu H.,  Bannister K.~W.,   et~al., 2019, MNRAS, 486, 166

\bibitem[\protect\citeauthoryear{Rafiei-Ravandi et~al.,}{Rafiei-Ravandi
  et~al.}{2021}]{rafiei21}
Rafiei-Ravandi M.,  et~al., 2021, arXiv preprint, 2106.04354

\bibitem[\protect\citeauthoryear{Ravi, Shannon  \& Jameson}{Ravi
  et~al.}{2015}]{ravi15}
Ravi V.,  Shannon R.~M.,   Jameson A.,  2015, ApJ Letters, 799, L5

\bibitem[\protect\citeauthoryear{Ravi et~al.,}{Ravi et~al.}{2016}]{ravi16}
Ravi V.,  et~al., 2016, Science, 354, 1249

\bibitem[\protect\citeauthoryear{Ridnaia et~al.,}{Ridnaia
  et~al.}{2021}]{ridnaia21}
Ridnaia A.,  et~al., 2021, Nature Astronomy, 5, 372

\bibitem[\protect\citeauthoryear{Schechter}{Schechter}{1976}]{schechter76}
Schechter P.,  1976, ApJ, 203, 297

\bibitem[\protect\citeauthoryear{Shannon et~al.,}{Shannon
  et~al.}{2018}]{shannon18}
Shannon R.~M.,  et~al., 2018, Nature, 562, 386

\bibitem[\protect\citeauthoryear{Shannon, Kumar  et~al.}{Shannon
  et~al.}{2019}]{shannon19}
Shannon R.~M.,  Kumar P.,   et~al., 2019, The Astronomer's Telegram, 12922

\bibitem[\protect\citeauthoryear{Spitler, Cordes  et~al.}{Spitler
  et~al.}{2014}]{spitler14}
Spitler L.~G.,  Cordes J.~M.,   et~al., 2014, ApJ, 790, 101

\bibitem[\protect\citeauthoryear{Tavani et~al.,}{Tavani
  et~al.}{2021}]{tavani21}
Tavani M.,  et~al., 2021, Nature Astronomy, 5, 401

\bibitem[\protect\citeauthoryear{Thornton et~al.,}{Thornton
  et~al.}{2013}]{thornton13}
Thornton D.,  et~al., 2013, Science, 341, 53

\bibitem[\protect\citeauthoryear{Zhang, Hobbs  et~al.}{Zhang
  et~al.}{2019}]{zhang19}
Zhang S.~B.,  Hobbs G.,   et~al., 2019, MNRAS: Letters, 484, L147

\bibitem[\protect\citeauthoryear{Zhang et~al.,}{Zhang et~al.}{2020}]{zhang20}
Zhang S.~B.,  et~al., 2020, ApJ Supplement Series, 249, 14

\bibitem[\protect\citeauthoryear{Zhang, Zhang  et~al.}{Zhang
  et~al.}{2021}]{2021MNRAS.501..157Z}
Zhang R.~C.,  Zhang B.,   et~al., 2021, MNRAS, 501, 157

\makeatother
\end{thebibliography}

\label{lastpage}

\end{document}